\documentclass[11pt, preprint]{aastex}
\usepackage{geometry}                
\usepackage{graphicx}
\usepackage{amssymb}
\usepackage{natbib}
\usepackage{amsmath}
\usepackage{epstopdf}
\newcommand{\eye}{\,{\bf I}}
\newcommand{\bx}{{\bf x}}
\newcommand{\bbox}{{\mathcal{B}}}
\newcommand{\disp}{{\xi}}
\newcommand{\by}{{\bf y}}
\newcommand{\bl}{{\bf L}}
\newcommand{\bc}{{\bf C}}

\newcommand{\bnabla}{{\boldsymbol \nabla}}
\newcommand{\bxi}{{\boldsymbol \xi}}
\newcommand{\bu}{{\boldsymbol u}}
\newcommand{\bb}{{\boldsymbol B}}
\newcommand{\bI}{{\boldsymbol I}}
\newcommand{\bo}{{\boldsymbol 0}}
\newcommand{\pts}{{\partial_{t_0}}}
\newcommand{\ptf}{{\partial_{t_1}}}

\begin{document}
\title{Propagation of seismic waves through a spatio-temporally fluctuating medium: Homogenization}
\author{Shravan M. Hanasoge\altaffilmark{1,2}, Laurent Gizon\altaffilmark{2,3} \& Guillaume Bal\altaffilmark{4}}
\altaffiltext{1}{Department of Geosciences, Princeton University, Princeton, NJ 08544, USA}
\altaffiltext{2}{Max-Planck-Institut f\"{u}r Sonnensystemforschung, 37191 Katlenburg-Lindau, Germany}
\altaffiltext{3}{Georg-August-Universit\"{a}t, Institut f\"{u}r Astrophysik, 37077 G\"{o}ttingen, Germany}
\altaffiltext{4}{Department of Applied and Physical Mathematics, Columbia University, New York 10027, USA}

\begin{abstract}
Measurements of seismic wave travel times at the photosphere of the Sun have enabled inferences of 
its interior structure and dynamics. 
In interpreting these measurements, the simplifying assumption that
waves propagate through a temporally stationary medium is almost universally invoked.
However, the Sun is in a constant state of evolution, on a broad range of spatio-temporal scales. 
At the zero wavelength limit, i.e., when the wavelength is much shorter than the scale over which the medium varies,
the WKBJ (ray) approximation may be applied. Here, we address
the other asymptotic end of the spectrum, the infinite wavelength limit, using the technique of homogenization.
We apply homogenization to scenarios where waves are propagating through rapidly varying media (spatially and temporally),
and derive effective models for the media. One consequence is that a scalar sound speed becomes a tensorial
wavespeed in the effective model and anisotropies can be induced depending on the nature of the perturbation. 
The second term in this asymptotic two-scale expansion, the so-called corrector, contains contributions due to higher-order scattering, leading
to the decoherence of the wavefield. This decoherence may be causally linked to the observed wave attenuation in the Sun.
Although the examples we consider here consist of periodic arrays of perturbations to the background, homogenization may be extended to ergodic and stationary random media.
This method may have broad implications for the manner in which we interpret seismic measurements in the Sun and for
modeling the effects of granulation on the scattering of waves and distortion of normal-mode eigenfunctions. 
\end{abstract}

\section{Introduction}
The Sun evolves continuously over a broad range of spatio-temporal scales.
Small-amplitude waves, which are stochastically excited by the action of vigorous near-surface turbulence, propagate through the
interior of the Sun and re-emerge at its surface. The Doppler shifting of spectral absorption lines, whose height of formation is altered by
these wave motions, provides a direct measurement of the seismic wavefield. Because we measure `noise', i.e., a superposition
of a multitude of randomly excited waves, we compute and average auto- and cross-correlations of the wavefield (between records at different spatial locations).
The wavefield of the Sun is described as an ergodic and temporally stationary random process, whose statistics are observed to be Gaussian \citep[e.g.,][]{woodard}. 

A critical goal of helioseismology is to image the properties of turbulence in the interior, the sub-surface structure of magnetic fields,  
and internal circulations of plasma. However, these phenomena that we want to image are undergoing constant evolution. The fundamental
question of how the properties of a temporally changing medium are imprinted in seismic measurements arises. 
The manner in which long wavelength waves couple with small-scale, temporally evolving granulation and whether this represents a strong
scattering regime and can therefore explain helioseismic wave damping \citep[e.g.,][]{duvall98} may be addressed using these techniques. 
Small-scale strong granulation flows are thought to distort normal-mode eigenfunctions \citep[e.g.,][]{brown84, murawski93, stix94, baldner12} 
and this may be investigated by deriving effective media and studying their anisotropy.

Homogenization, a mathematical technique based on a two-scale asymptotic analysis, is a treatment of wave propagation in the long-wavelength
limit. 
It gives us a means for deriving an effective medium, replacing a rapidly varying medium by a smoother equivalent that to first order is able
to model wave propagation on large wavelengths.
An analogy may be drawn to Floquet-Bloch descriptions \citep{floquet, bloch}, which addresses the energy eigenfunction of an electron within a periodic potential. 
The theory of homogenization, easily realized for wave propagation in deterministic periodic media, was extended by \citet{kozlov79} and \citet{varadhan82} to ergodic random media by allowing
for the periodicity length scale to become infinite. In other words, \citet{kozlov79} and \citet{varadhan82} realized that an ergodic random medium is essentially a periodic medium, but repeating on infinitely long length scales.
Thus the method assists us in interpreting the (mostly) horizontally stationary and ergodic random process
that helioseismic waves are described by \citep[e.g.,][]{woodard}.

In this article, we choose to study the problem of wave propagation amid a spatio-temporally periodically fluctuating array speed fluctuations in the context of a much simplified wave equation. For the theory of homogenization, we refer the reader to, e.g., \citet{lions78}. Homogenization theory, originally developed for periodic structures, also applies in a variety of random media, under the assumption that their statistics are translationally invariant and ergodic in an appropriate sense; see \citet{JKO-SV-94}.
For now it is instructive to deal with the periodic case.

\section{The wave equations}
The propagation of linear small-amplitude waves in a spatio-temporally evolving medium with flows and magnetic fields 
is fairly complicated. A variational treatment of the governing equations and the derivation of the formidable set
of full equations may be found in \citet{webb05}. We merely reproduce them here.

Defining the material derivative of the wave displacement by $\dot\bxi$, 
\begin{equation}
\dot\bxi = \frac{\partial\bxi}{\partial t} + \bu\cdot\bnabla\bxi,
\end{equation}
the oscillations are described by
\begin{eqnarray}
\rho\partial_t{\dot \bxi} &=& -\rho\bu\cdot\bnabla\dot\bxi +\bnabla\cdot[ (\rho c^2 - p)\bnabla\cdot\bxi\,\,\bI + \left(p + B^2/2\right)(\bnabla\bxi)^t\nonumber\\
&&-(\bb\bb:\bnabla\bxi - B^2/2\bnabla\cdot\bxi)\bI  - \bb(\bb\bnabla\cdot\bxi - \bb\cdot\bnabla\bxi)  ] - \rho\bxi\cdot\bnabla\bnabla\phi,\label{gov.eq}
\end{eqnarray}
where $t$ is time, $\bx$ is space, $\bnabla$ is the covariant spatial derivative, $\bI$ is the dyadic identity tensor,
$p(\bx,t)$ background pressure, $\rho(\bx,t)$ background density, $c(\bx,t)$ sound speed, $\bb(\bx,t)$ background magnetic field, $\bu(\bx,t)$ background flows,
gravity ${\bf g} = -\bnabla\phi$ and $S(\bx,t)$ background entropy.
 The following tensor notation applies $[\bnabla\bxi]_{ij} = \partial_i\xi_j$ and
$[(\bnabla\bxi)^t]_{ij} = \partial_j\xi_i$, where $\partial_i \equiv \partial/\partial x^i$ denotes the covariant spatial derivative.

\section{Temporal Homogenization}
We consider a medium where the sound speed and density are
periodically fluctuating in time on a timescale much shorter than the period of the waves of interest. For simplicity's sake, there will be no
background flows or magnetic fields, i.e., equation~(\ref{gov.eq}) with $\bb = \bo$, $\bu = \bo$, no entropy waves, $\Delta S = 0$ and we
assume constant background pressure, i.e., $\bnabla p = \bo$.
We consider a periodically varying soundspeed $c(\bx,t)$ such that $c(\bx,t+T) = c(\bx,t)$, 
where $T$ is the periodicity timescale, much smaller than the dominant wave period, and $t$ is time.
The differential equation of interest is
\begin{equation}
\rho \partial_t^2 \disp - (\rho \partial_t\ln c)\, \partial_t\disp - \rho c^2\nabla^2 \disp = 0,\label{eq.orig}
\end{equation}
where $\disp$ is the scalar wavefield displacement. Temporally fluctuating coefficients can pump energy into the wave system. However, this particular form of the wave equation conserves
energy. To demonstrate that this is the case, we divide equation~(\ref{eq.orig}) by $c^2$, multiply it by $\partial_t\disp$ and integrate over volume to obtain
\begin{equation}
\int d\bx\, \left(\frac{1}{c^2}\partial_t^2\disp - \frac{1}{c^3}\partial_t c\, \partial_t\disp - \nabla^2\disp\right) \partial_t\disp = 0,
\end{equation}
which may be manipulated further,
\begin{equation}
\int d\bx\,\left[ \partial_t\left(\frac{1}{c^2}\frac{\dot\disp^2}{2}\right) + \partial_t \frac{||\bnabla\disp||^2}{2} \right] = 0,\label{ener1}
\end{equation}
where $\dot\disp = \partial_t\disp$ and we have assumed, for the sake of simplicity, that the boundaries are periodic. This allows
us to drop boundary-related integrals when using Gauss's theorem to transform the spatial gradient term into the form~(\ref{ener1}).
Defining the wave energy as
\begin{equation}
{\mathcal E} = \int d\bx\,\left\{ \frac{1}{c^2}\frac{\dot\disp^2}{2}+ \frac{||\bnabla\disp||^2}{2}\right\}
\end{equation}
equation~(\ref{ener1}) indicates that it is an invariant, i.e.,
\begin{equation}
\partial_t {\mathcal E} =0.
\end{equation}
Given the existence of the energy invariant, we are assured that classical techniques of homogenization are applicable.
To facilitate the means of retrieving the homogenized solution, we rewrite the wave equation~(\ref{eq.orig}) thus
\begin{equation}
\partial_t\left(\frac{\partial_t\disp}{c}\right) - c\nabla^2\disp = 0.
\end{equation}
Introducing two temporal scales, $t_0$ and $t_1$, the fast and slow scales respectively, and
a small parameter $\varepsilon = \nu_0 T \ll 1$, where $\nu_0$ is the dominant wave frequency,
we note that the first and second time derivatives are
\begin{eqnarray}
&&\partial_t = \pts + \frac{1}{\varepsilon}\ptf\\
&&\partial^2_t = \pts^2 + \frac{2}{\varepsilon}\ptf\pts + \frac{1}{\varepsilon^2}\ptf^2.
\end{eqnarray}
We expand $u$ into the series $\disp = \disp_0 + \varepsilon \disp_1 + \varepsilon^2 \disp_2 + O(\varepsilon^3)$, where
$\disp_i = \disp_i(\bx, t_0, t_1)$ and $\disp_i(\bx, t_0, t_1+T) = \disp_i(\bx, t_0, t_1)$.
At order $\varepsilon^{-2}$, the equation reads,
\begin{equation}
\ptf\left(\frac{\ptf\disp_0}{c}\right) = 0.
\end{equation}
Multiplying by $\ptf\disp_0$, integrating over the fast temporal scale, integrating by parts and invoking periodicity, we obtain
\begin{equation}
\int dt_1 \,\ptf\disp_0\,\ptf\left(\frac{\ptf\disp_0}{c}\right) = -\int dt_1\,  \frac{1}{c}(\ptf \disp_0)^2=0,
\end{equation}
which is negative definite integral unless $\ptf\disp_0 \equiv 0$, forcing $\disp_0$ to be only a function of $t_0$, i.e., $\disp_0 = \disp_0(\bx,t_0)$.
At order $\varepsilon^{-1}$, the expansion provides
\begin{equation}
\pts\left( \frac{\ptf\disp_0}{c}\right) + \ptf\left( \frac{\pts\disp_0 + \ptf\disp_1}{c}\right) = 0,
\end{equation}
or,
\begin{equation}
\ptf\left( \frac{\pts\disp_0 + \ptf\disp_1}{c}\right) = 0.
\end{equation}
Invoking the ansatz $\ptf\disp_1 = h(t_1)\pts\disp_0$, together
with the constraint arising from periodicity that
\begin{equation}
\int_0^T dt_1\, h(t_1) = 0,
\end{equation}
we have
\begin{equation}
h(t_1) = \frac{c}{\langle c \rangle} - 1,
\end{equation}
where 
\begin{equation}
\langle c\rangle = \frac{1}{T}\int_0^T dt_1\, c.\label{c2hom}
\end{equation}
Thus the following relationship holds
\begin{equation}
 \frac{\pts\disp_0 + \ptf\disp_1}{c} = \frac{1}{\langle c\rangle}\pts\disp_0,\label{rel.ref}
\end{equation}
and the corrector $\disp_1$ is given by
\begin{equation}
\disp_1 = \left(\frac{1}{\langle c \rangle}\int_0^t dt_1\, c - t\right) \pts\disp_0.
\end{equation}
 At order $\varepsilon^0$, we obtain
\begin{equation}
\pts\left(\frac{\pts\disp_0 + \ptf\disp_1}{c}\right) + \ptf\left(\frac{\pts\disp_1 + \ptf\disp_2}{c}\right) - c\nabla^2 \disp_0 = 0,
\end{equation}
and integrating over the fast temporal scale, invoking periodicity and using relationship~(\ref{rel.ref}), we obtain
the homogenized differential equation
\begin{equation}
\pts^2\disp_0 - \langle c \rangle^2\nabla^2\disp_0  = 0.
\end{equation}

As we will see in subsequent sections, the bulk modulus $\rho c^2$ is typically
replaced by a tensorial wave speed that varies as a function of the propagation direction. The effective
equation in general acquires greater complexity than the original form.
It is also important to note that this equation does not always produce a stable solution, and that oscillating
coefficients can destabilize the wave equation \citep[see, e.g.,][]{colombini}. This occurs because oscillating
coefficients pump energy into the system and in this particular scenario, we control this by introducing a damping
term.

\section{Numerical tests}
\subsection{Sound-speed perturbation}
\begin{figure}[!ht]
\begin{centering}
\epsscale{1}
\plotone{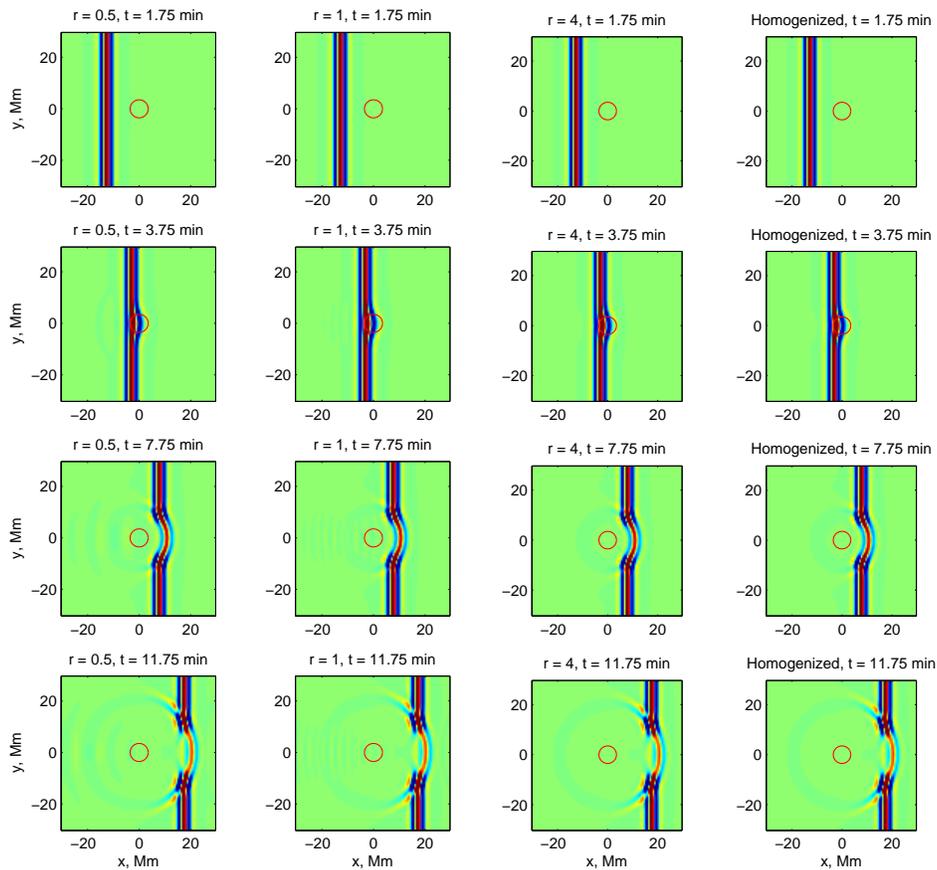}
\caption{Snapshots of the wavefield at four time instants (top to bottom), for different ratios $r = \omega/\omega_0$, where 
$\omega_0$ is the central frequency of the wavepacket and $\omega$ is the oscillation frequency of the sound-speed perturbation~(\ref{pert}).
The homogenized solution, i.e., with perturbation~(\ref{homogpert}), is shown on the fourth column. The circle in the center marks the
location of the FWHM of the sound-speed perturbation.
It is seen that the wavefields for all $r$ are very similar to the homogenized solution with the $r=1$ case showing the largest discernible difference. 
\label{snapshot}}
\end{centering}
\end{figure}

\begin{figure}[!ht]
\begin{centering}
\epsscale{1}
\plotone{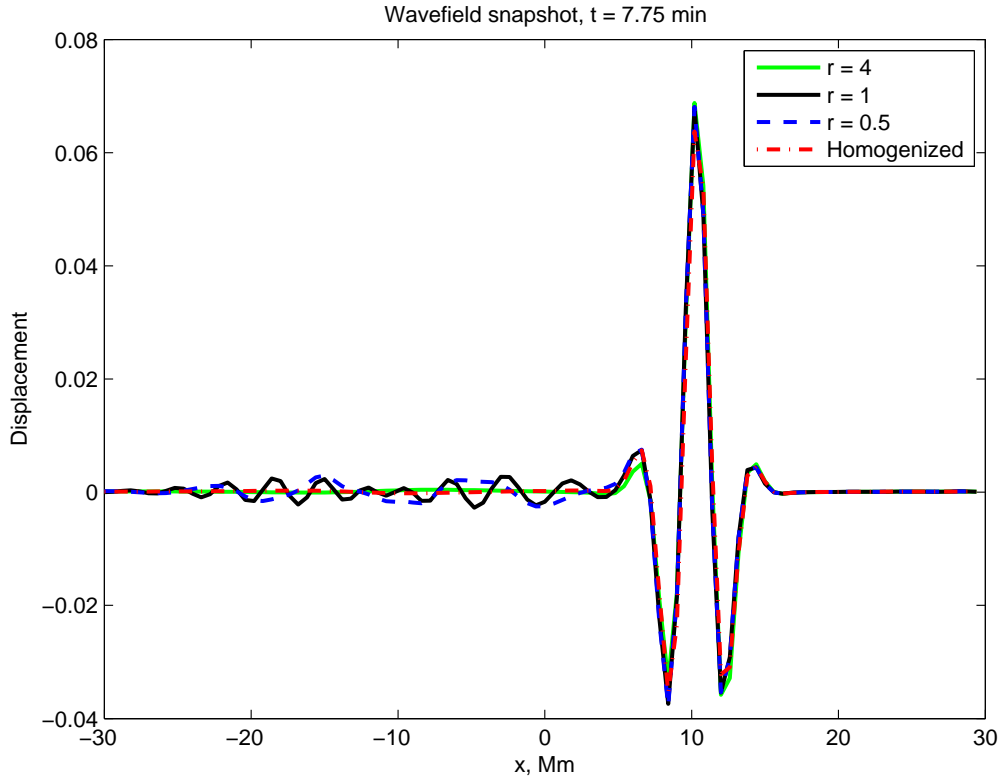}
\caption{A cut through $y = 0$~Mm of the wave field for different values of the ratio $r = \omega/\omega_0$. It is seen
that the wave is most strongly scattered when the sound-speed perturbation oscillates at the same frequency as the wave,
suggesting some form of resonant interaction.
The $r=4$ line is almost indistinguishable from the homogenized solution.
\label{cutime}}
\end{centering}
\end{figure}

We study equation~(\ref{eq.orig}) numerically in order to characterize the bounds and effectiveness of homogenization 
at addressing wavespeed perturbations. 
We solve the equation using a pseudo-spectral solver, computing horizontal derivatives using a fast Fourier transform and
evolving it in time through the repeated application of an optimized
five-stage second-order Runge-Kutta scheme \citep{hu}. The horizontal
boundaries are periodic. We set off a one-way Gaussian wave packet at t=0, of central
wavelength 3.33 Mm, central frequency of $\omega_0/(2\pi) = 3$ mHz, and with a  nominal sound speed of $c_0 = 10$ km/s.
The  full width at half maximum (FWHM) of this Gaussian wavepacket is 10 Mm.
The background medium contains a Gaussian ball shaped sound-speed perturbation, whose
FWHM is one wavelength, i.e., 3.33 Mm. The amplitude of the perturbation oscillates in time,
varying from 0 to 40\% in sound speed. The form of this perturbation is described by
\begin{equation}
c = c_0\left[1 + A\,\exp\left(- \frac{x^2 + y^2}{\sigma^2}\right)\,\sin^2\left(\omega t\right) \right],\label{pert}
\end{equation}
where $c_0 = 10$ km/s, $A = 2$ and $\sigma = 3.33$ Mm. It is seen from equation~(\ref{c2hom})
that, when $\omega \gg \omega_0$, the homogenized, effective sound-speed squared is given by
\begin{equation}
\langle c\rangle = c_0\left[1 + \frac{A}{2}\,\exp\left(- \frac{x^2 + y^2}{\sigma^2}\right)\,\right].\label{homogpert}
\end{equation}
In Figure~\ref{snapshot}, we show the solution at four different instants for four different cases,
where $r = \omega/\omega_0 = [0.5, 1.0, 2.0]$ and the homogenized, time-stationary sound-speed solution.
Figure~\ref{cutime} displays a cut of the wavefield along the centerline of the $x$-axis. The homogenized
solution together with the corrector is compared with the full solution, and it is seen that the performance is worst for the $r=0.5, 1$ cases, where some form of
temporal scattering resonance may be occurring.

\section{Spatial Homogenization}
We now consider a medium where the sound speed and density are temporally stationary but spatially
fluctuate periodically on a length scale much shorter than the wavelengths of interest. 
We consider a periodically varying soundspeed $c(\bx,\bx/\varepsilon)$ and density 
$\rho(\bx,\bx/\varepsilon)$, where box $\bbox = [0, L_1)\times[0,L_2)$ describes
the unit periodic box (much smaller than the dominant wavelength) that is used to tile the entire domain, 
$\bx$ is the `slow' coordinate and $\by = \bx/\varepsilon$ the fast coordinate. The differential equation of interest is
\begin{equation}\label{eq:wavexi}
\rho\partial_t^2 \disp - \bnabla\cdot(\rho c^2\bnabla \disp) = 0,
\end{equation}
where $\disp$ is the wavefield and $\bnabla$ is the covariant spatial derivative. 
It may be verified that this differential equation possesses an energy invariant, given by
\begin{equation}
{\mathcal{E}} = \int d\bx\,\left\{\frac{\rho \dot\disp^2}{2} + \rho c^2\frac{||\bnabla\disp||^2}{2}\right\},
\end{equation}
i.e., $\partial_t{\mathcal E}=0$. The existence of an energy invariant guarantees the convergence of a homogenization expansion.
The two-scale representation of the spatial derivative is given by
\begin{equation}
\bnabla = \bnabla_\bx + \varepsilon^{-1}\bnabla_\by.
\end{equation}
Introducing the ansatz $\disp = \disp_0(\bx, \bx/\varepsilon,t) + \varepsilon \disp_1(\bx,\bx/\varepsilon,t) + \varepsilon^2 \disp_2(\bx,\bx/\varepsilon,t) + ...$  Collecting terms of $O(\varepsilon^{-2})$, we have
\begin{equation}
\bnabla_\by( \rho c^2 \bnabla \disp_0) = 0,
\end{equation}
which, invoking the periodicity over $\bl$, may be manipulated as follows
\begin{equation}
\int_{\bbox} d\by\, \disp_0\bnabla_\by\cdot( \rho c^2 \bnabla \disp_0) = -\int_\bbox d\by\,  \rho c^2 ||\bnabla \disp_0||^2 = 0,
\end{equation}
implying that $\bnabla_\by \disp_0 = {\bf 0}$ or $\disp_0 \equiv \disp_0(\bx,t)$. The following definition applies
\begin{equation}
\int_\bbox d\by = \int_0^{L_1} dy_1 \int_0^{L_2} dy_2,
\end{equation}
where $\by = (y_1, y_2)$.
This is in line with expectation, since, at leading order,
the solution is presumably dominated by effects on the scale of the wavelength.
 At order $\varepsilon^{-1}$, we obtain
\begin{equation}
\bnabla_\by\cdot( \rho c^2 \bnabla_\by \disp_1) + \bnabla_\bx\cdot( \rho c^2 \bnabla_\by \disp_0) + \bnabla_\by\cdot( \rho c^2 \bnabla_\bx \disp_0) = 0.
\end{equation}
Invoking the result that $\disp_0 = \disp_0(\bx,t)$, this simplifies to
\begin{equation}
\bnabla_\by\cdot[ \rho c^2 (\bnabla_\by \disp_1 + \bnabla_\bx \disp_0)] =  0.
\end{equation}
We seek solutions of the form \citep[e.g.,][]{lions78} $ \disp_1 = {\bf h}(\by)\cdot\bnabla_\bx \disp_0$,
which implies
\begin{equation}
\bnabla_\by\cdot[\rho c^2(\bnabla_\by{\bf h} + \eye)]\cdot\bnabla_\bx \disp_0 = 0,
\end{equation}
where $\eye$ is identity tensor.
This produces the following elliptic equation
\begin{equation}
\bnabla_\by\cdot[\rho c^2(\bnabla_\by{\bf h} + \eye)] = {\bf 0},\label{elip}
\end{equation}
whose solution gives us the the corrector $\disp_1 = {\bf h}\cdot\bnabla \disp_0$. This is the classical {\it cell problem} in homogenization. 
Note that this is an implicit equation that does not, in general, possess a closed form or explicit solution.
At order $\varepsilon^0$, we obtain
\begin{equation}
\rho\partial_t^2 \disp_0 - \bnabla_\bx\cdot[\rho c^2(\bnabla_\bx \disp_0 + \bnabla_\by \disp_1)] -\bnabla_\by\cdot(\rho c^2\bnabla_\by \disp_2) = 0,
\end{equation}
or 
\begin{equation}
\rho\partial_t^2 \disp_0 - \bnabla_\bx\cdot[\rho c^2(\bnabla_\by{\bf h} + \eye)\cdot\bnabla_\bx \disp_0] -\bnabla_\by\cdot(\rho c^2\bnabla_\by \disp_2) = 0.
\end{equation}
Integrating over the fast variable and invoking periodicity,
\begin{equation}
\rho_*\partial_t^2 \disp_0 - \bnabla_\bx\cdot[\bc_*\cdot\bnabla_\bx \disp_0]  = 0,
\end{equation}
where the following definitions hold
\begin{eqnarray}
\rho_* &=& \frac{1}{L_1 L_2}\int_{\bbox} d\by\, \rho(\bx,\by),\label{eq:bstar}\\
\bc_* &=& \frac{1}{L_1 L_2}\int_{\bbox} d\by\, \rho c^2(\bnabla_\by{\bf h} + \eye).\label{tensorwave}
\end{eqnarray}
As a consequence of spatial homogenization, the simple scalar wavespeed $\rho c^2$ has now been transformed to wavespeed tensor $\bc$ whose
description is obtained by solving the implicit partial differential equation~(\ref{elip}).
Evidently, the case of spatial homogenization is substantially more complicated than the temporal analog.

More complicated yet and in fact much less understood mathematically is the practically interesting case of fluctuations in both time and space. 
It is known for some specific choices of the coefficients $\rho$ and $c^2$ that the wave equation may not have bounded solutions \citep[e.g.,][]{colombini}. 
Again, however, with the existence of an energy invariant, the multi-scale expansion is convergent,
in which case a dual spatio-temporal homogenization procedure may be applied.
See \citet{lions78} for additional details on the homogenization of the wave equation in periodic media.

\subsection{Numerical example}
\begin{figure}[!ht]
\begin{centering}
\epsscale{1}
\plotone{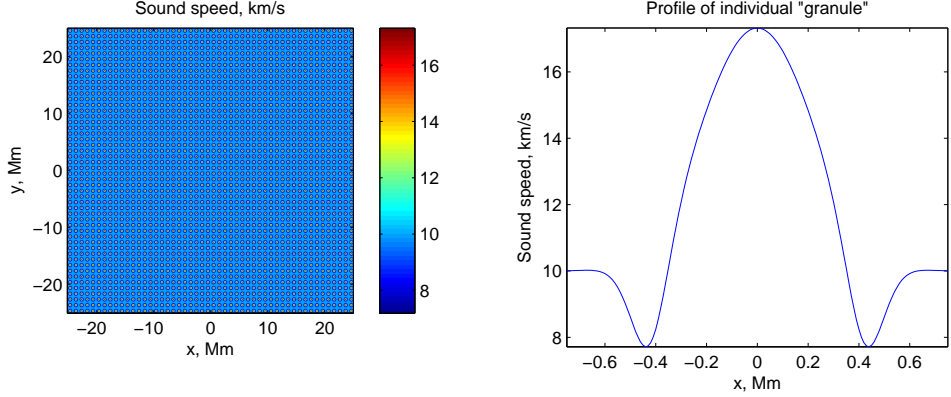}
\caption{Periodic grid of sound-speed perturbations used in our simulations (left panel) and the
profile of an individual perturbation (right panel). The nominal sound speed is 10 km/s and the perturbation
is positive in the center, i.e. an increased sound speed, surrounded by a annulus of decreased speed.  This choice is made to mimic the
hot central core of granules and cold surrounding down flows. The effective size of each granular cell is around 1 Mm, and the wavelength of
the incoming wave is 3.34 Mm.
The sound speed is seen to vary by over a factor of two.
\label{soundspeed}}
\end{centering}
\end{figure}

Granulation in the Sun is a process that is spatially horizontally `periodic' and substantially smaller than typical acoustic wavelengths. In this simplistic model of
wave propagation through granules, we consider a spatially periodic grid of sound speed perturbations, shown in Figure~\ref{soundspeed}. 
The elliptic partial differential equation~(\ref{elip}) is solved by relaxing it to a steady state. The two components of that vector equation are
\begin{eqnarray}
\bnabla\cdot[\rho c^2(\bnabla h_x + {\bf e}_x)] &=& 0,\nonumber\\
\bnabla\cdot[\rho c^2(\bnabla h_y + {\bf e}_y)] &=& 0,
\end{eqnarray}
where $h_x$ and $h_y$ are the $x$ and $y$ components of vector ${\bf h}$ and  ${\bf e}_x,~{\bf e}_y$ are the unit vectors along the $x$ and $y$ axes.
The relaxation equation for the $h_x$ equation is given by
\begin{equation}
\partial_t h_x = \bnabla\cdot[\rho c^2(\bnabla h_x + {\bf e}_x)],\label{relax}
\end{equation}
where $h_x$ now is a function of time and space.
This diffusion equation eventually relaxes
to the steady-state $\bnabla\cdot[\rho c^2(\bnabla h_x + {\bf e}_x)] = 0$. We temporally evolve~(\ref{relax}) till a steady state is achieved. 
The code is 
tested against an analytical solution that is known for perturbations that are only functions of one coordinate, e.g., where $c^2 = c^2(x)$ only. In such
a case, the wave speed tensor ${\bf C}_*$ has diagonal components, given by the harmonic and simple means.

Once we obtain the vector ${\bf h}$, we compute the wavespeed tensor using equation~(\ref{tensorwave}).
The cell problem studied here has cylindrically symmetry, i.e., each of the sound-speed perturbations is azimuthally symmetric around its axis. This
implies that the wave speed tensor will contain identical components in the $xx$ and $yy$ planes, providing us another way to test the code.

\begin{figure}[!ht]
\begin{centering}
\epsscale{1}
\plotone{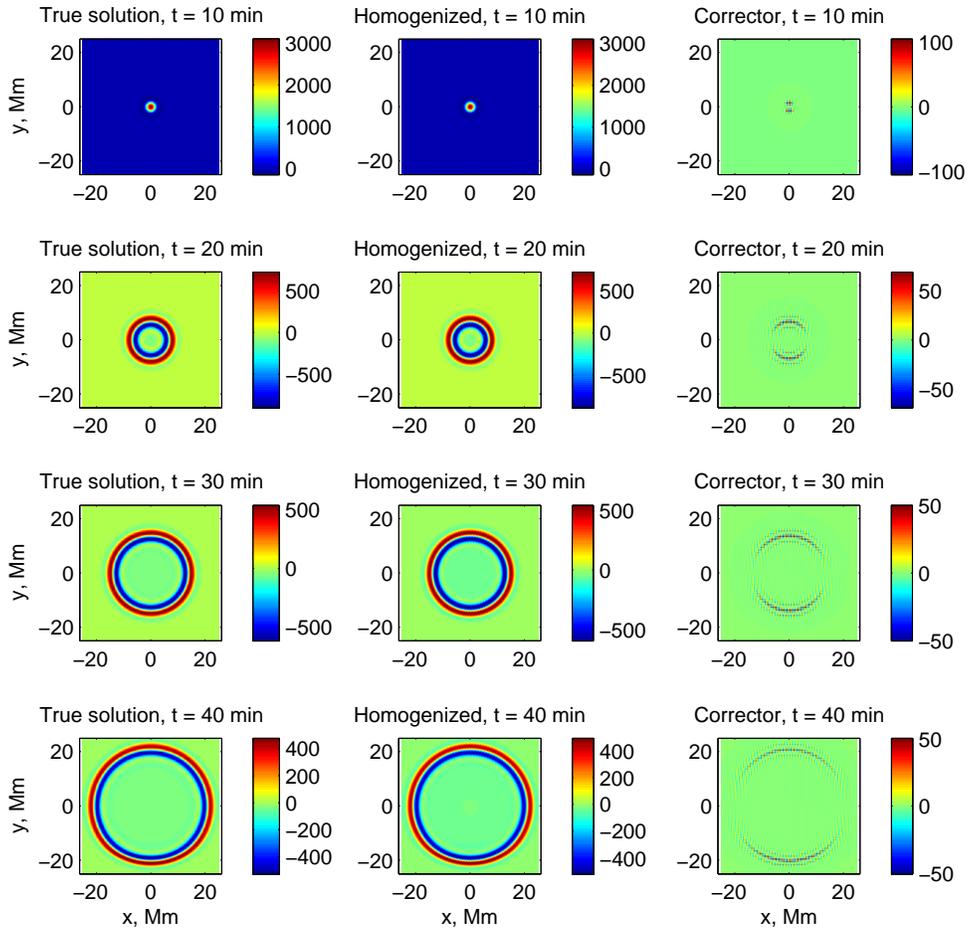}\vspace{-1cm}
\caption{Snapshots of waves propagating through the sound-speed perturbations of Figure~\ref{soundspeed} (left column) and 
through a homogenized model (middle column). The wave displacements are for all practical purposes identical and are therefore not
shown here. Homogenization succeeds in accurately capturing the wavefield in the asymptotic infinite-wavelength limit. The corrector,
given by $\xi_1 = {\bf h}\cdot\bnabla_\bx \xi_0$, is shown on the third column, and shows the higher-order scattering term.
\label{snapspace}}
\end{centering}
\end{figure}

Once we obtain the homogenized wave speed tensor, which is essentially a constant sound speed in the $xx$ and $yy$ planes, we can compare the 
true and homogenized solutions.
We set off a source at the center of the computational domain and compare the homogenized and true solutions in Figure~\ref{snapspace}.
The corrector, given by $\xi_1 = {\bf h}\cdot\bnabla\xi_0$, contains the higher-order scattering terms not fully captured by the homogenized
solution. We show the corrector also in Figure~\ref{snapspace}, which contribute to a decoherence of the input wave packet, thereby contributing
to observed wave attenuation. 
The wave fields are practically identical, indicating that the infinite wavelength limit works very accurately at modeling these `granules'. 

\begin{figure}[!ht]
\begin{centering}
\epsscale{0.75}
\plotone{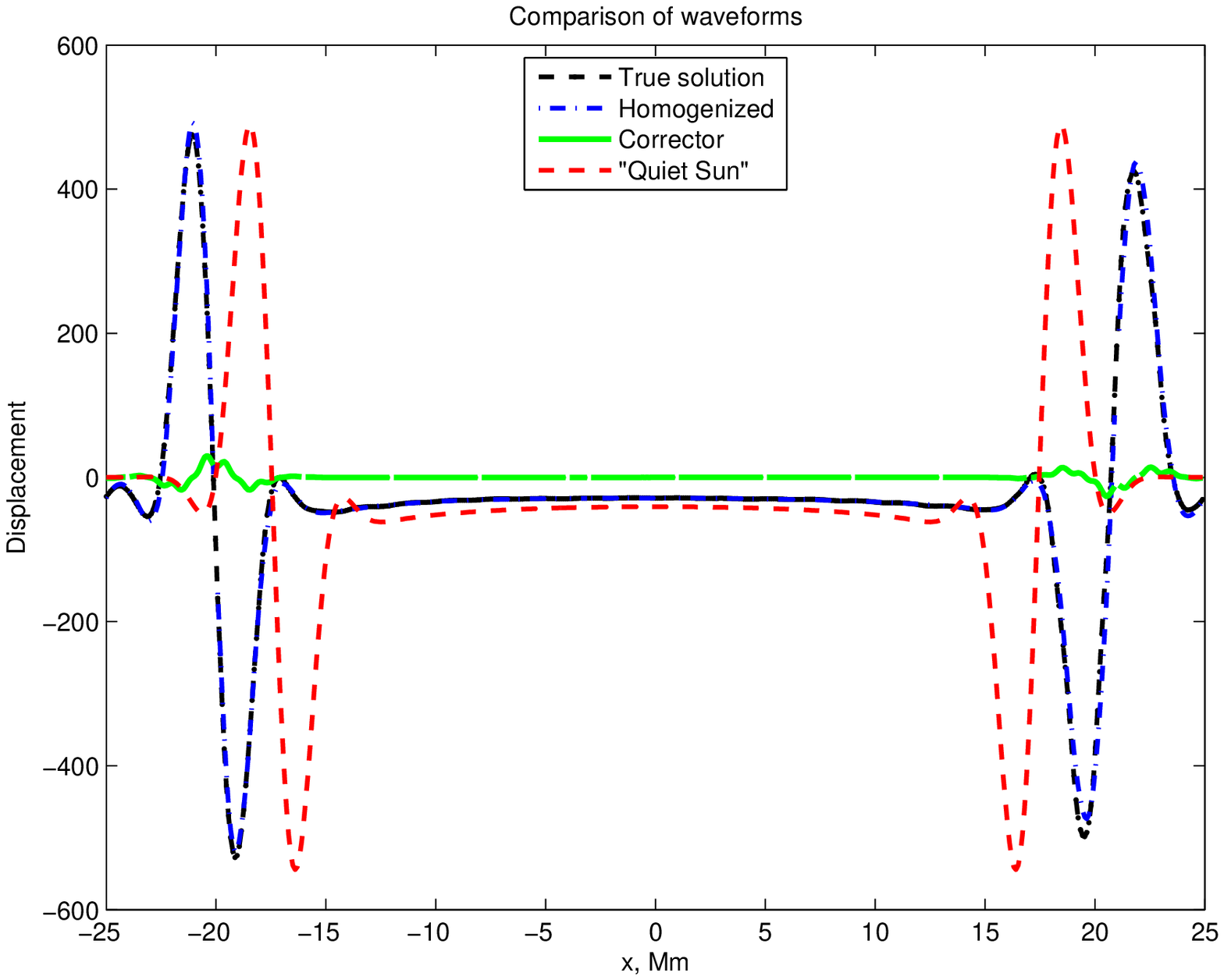}
\caption{Cut along $y=0$ of the true, homogenized, and ``quiet" wavefields at $t = 40$ min, where the nominal $c = 10$ km/s sound speed is used in the quiet calculation.
The homogenized and true solutions are indistinguishable while a systematic time shift is seen between the quiet and true solutions, demonstrating that the dominant impact of the ``granules" on the wavefield is to induce phase shifts. The corrector, the thick green line, contains the higher-order scattering term not captured by the homogenized solution.
\label{quiettruecomp}}
\end{centering}
\end{figure}

\section{Random Media}
In the cases we have considered thus far, the media consists of periodic arrays of scatterers. It was shown by \citet{varadhan82}
that random media where the time-independent perturbations have short correlation length scales and 
are drawn from stationary and ergodic distributions, the expectation value of the wavefield
is given by the ensemble average of equation~(\ref{tensorwave}). The result rests on the argument that random media are the limiting case of
a period medium with an infinite periodicity length scale. See \cite{JKO-SV-94} for additional details on the theory of homogenization in random media. Although mathematically more difficult to establish than homogenization theory in periodic media, the main conclusions drawn in periodic media typically also hold in random media. Indeed, seeing random media as a limit of periodic media with increasing cell size, we may obtain the homogenized coefficients in random media $\rho_*$ and $\bc_*$ as the limits given in \eqref{eq:bstar}-\eqref{tensorwave} as the sizes $L_1$ and $L_2$ tend to $\infty$.

\begin{figure}[!ht]
\begin{centering}
\epsscale{1}
\plotone{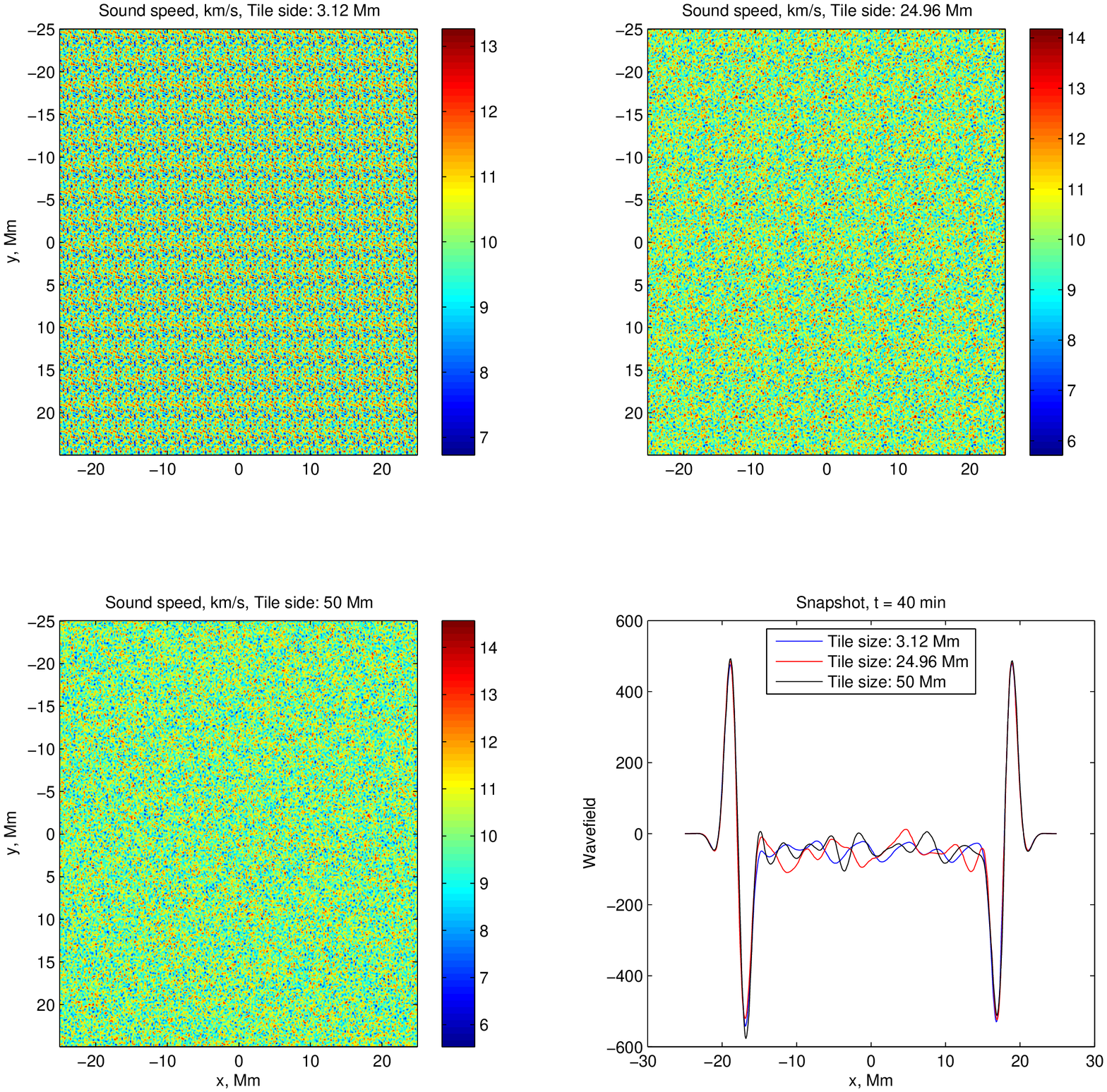}
\caption{Wave propagation in random media. Sound speed perturbations are generated by first creating a unit tile and filling the computational
domain with periodic repetitions of the tile. Random media may be thought of as a periodic array of perturbations, 
but with an infinite periodicity length scale, and so \citet{varadhan82} demonstrated
that there exists an effective medium that describes wave propagation through stationary and ergodic fields of perturbations. 
The unit tile is successively increased in size till the entire domain domain is covered by one
square (i.e., the poor man's approximation to an ``infinite" periodicity length scale). The wavefield, to leading order, is identical in all the cases.
\label{random.result}}
\end{centering}
\end{figure}

In this section, we show a simple case where we tile the 2-D computational domain
with a randomly generated square of sound-speed perturbations. The perturbations are drawn from a zero-mean uniform distribution with an amplitude
of 8 km/s, where the nominal sound speed is 10 km/s.
This square is successively increased in size till it is the size of the entire domain. In other words, for a
computational domain of $512\times 512$ we choose tiles of sizes $32\times32, 64\times64, 128\times128, 256\times256$ and $512\times512$.
The full medium is then filtered to remove the top third highest spatial frequencies (up to the spatial Nyquist) in order to prevent aliasing from corrupting the numerical simulation
\citep[Orszag's two-thirds rule][]{orszag71}. The dominant power in the spectrum of the fluctuations is on length scales smaller than the peak
wavelength of the wave, 3.33 Mm.
To leading order, we show that the wavefield is the same in all the cases in Figure~\ref{random.result}. A more thorough investigation
may be performed, where for a given size of the cell, the cell problem~(\ref{elip}) is solved for a number of realizations and the variance
of the homogenized coefficient~(\ref{tensorwave}) is estimated. It can be shown that the variance of the homogenized
coefficient falls as $N^{-d}$, where $N=L_1=L_2$ is the size of the tile and $d$ describes some rate. 

Optimal rates of convergence for random coefficients with short-range correlations may be obtained (for a slightly modified problem) in \citet{GO-AP-11}. These analyses are difficult and not known for large classes of processes. 

\section{Conclusions}
The WKBJ approximation represents the zero wavelength limit of wave propagation, where
the scale over which the structure changes is substantially larger than the wavelength. However,
the Sun displays structure over a broad range of scales and consequently, the asymptotic infinite wavelength limit
is also very important to understand.

It is believed small-scale granulation likely plays a critical role in scattering waves and distorting eigenfunctions
of normal modes \citep{brown84, baldner12}.  From the numerical experiments we have performed here, we find that
when spatio-temporal scales are separated, scattering will be very weak and that strong scattering happens
only in the case where the scales overlap. This may have important implications for granular scattering of waves,
where the spatial scale separation is significant but the temporal scales of its evolution are similar. 
Not surprisingly, asymptotic methods break down in this strong scattering regime.
Modeling these effects is an important step towards interpreting seismic measurements appropriately.
Most interestingly the analysis reveals that the effective medium possesses a tensorial wavespeed and can potentially
induce anisotropy in wave propagation. We compute the corrector, which represents higher-order scattering that contributes to the overall decoherence of the
wavefield. 

A powerful extension of the periodic case (that we have studied here) is to ergodic random media where the probability density function
describing the randomness is translationally invariant. Granulation and supergranulation fall into this regime, both being described by translationally
horizontally invariant quasi-random processes.

\acknowledgements
S. M. H. is funded by NASA grant NNX11AB63G. This work is an effort to understand cross correlations in 
helioseismology in the context of DFG SFB 963 `Astrophysical Flow Instabilities and Turbulence' (Project A1). 
S. M. H. and L. G. would like to thank the Courant Institute, New York University for their hospitality.

\bibliographystyle{apj}

\end{document}